%% file: Paper.tex
\documentclass[conference]{IEEEtran}
\IEEEoverridecommandlockouts
\usepackage{balance}
\usepackage{cite}
\usepackage{amsmath,amssymb,amsfonts}
\input{def}

\usetikzlibrary{external} % Load the TikZ external library
\tikzset{external/optimize=false}
\tikzset{external/system call={pdflatex \tikzexternalcheckshellescape -halt-on-error -interaction=batchmode -jobname "\image" "\texsource" && epstopdf "\image".pdf --outfile="\image".eps}}

\begin{document}

\title{
Joint Access Point Selection and Precoder Design under Statistical CSI
}
\author{\IEEEauthorblockN{Tim N. Faisst, Franz Weißer, and Wolfgang Utschick\\}
\IEEEauthorblockA{\textit{TUM School of Computation, Information and Technology, Technical University of Munich, Germany} \\
\{timniklas.faisst, franz.weisser, utschick\}@tum.de}
}

\maketitle

\begin{abstract}
    This work addresses joint \ac{AP} selection and precoding for sum-rate maximization under statistical \ac{CSI} in multi-\ac{AP} multi-user systems. 
    To this end, we propose two approaches. 
    The first method is an iterative alternating optimization algorithm that updates the precoding vectors via the \ac{SWMMSE} algorithm and the assignment variables via a projected gradient descent step. 
    The second method is a \ac{GNN}-based framework that solves the same problem in a single forward pass during inference.  
    Building on an attention-based Edge-\ac{GNN} architecture, we extend it to a multi-\ac{AP} scenario, enabling the joint learning of assignment variables and precoding vectors from statistical \ac{CSI} alone.
    Results show that the \ac{GNN} outperforms the iterative algorithm across the tested \ac{SNR} range and generalizes to varying numbers of users with comparable performance. 
    Both approaches are also compared to various baseline techniques.
\end{abstract}

\begin{IEEEkeywords}
	Statistical CSI, access point selection, precoding, sum-rate maximization, stochastic WMMSE, graph neural networks
\end{IEEEkeywords}

\input{intro}

\input{system}
\input{method}
\input{simulations}

\input{conclusion}

%\newpage
\balance
\bibliographystyle{IEEEtran}
  
%\bibliography{...}
\bibliography{IEEEabrv,mybib}

\end{document}

%% file: def.tex
\usepackage{array}
\usepackage[caption=false,font=footnotesize,position=bottom]{subfig}
\usepackage{capt-of}
\usepackage{stfloats}
\usepackage{url}
\usepackage{graphicx}
\usepackage{textcomp}
\usepackage{xcolor}
\usepackage{tikz}
\usepackage{ellipsis}
\usetikzlibrary{calc}
\usetikzlibrary{decorations.pathreplacing,decorations.markings,shapes.geometric}
\usetikzlibrary{calc,patterns,angles,quotes,shapes,arrows.meta}
\usetikzlibrary{chains}
\usepackage{balance}
\usepackage{cite}
\usepackage{amsmath,amssymb,amsfonts}
\usepackage{algorithm}
\usepackage{algpseudocode}
\usepackage{blindtext}
\usepackage[capitalise]{cleveref}

\usepackage[utf8]{inputenc}

\usepackage{mathtools}
\usepackage{bm}
\usepackage{etoolbox}
\usepackage{scalerel}

\usepackage{fancyhdr}

\newcommand{\va}{{\bm{a}}}
\newcommand{\vb}{{\bm{b}}}
\newcommand{\vc}{{\bm{c}}}

\newcommand{\vf}{{\bm{f}}}

\newcommand{\vh}{{\bm{h}}}
\newcommand{\vk}{{\bm{k}}}

\newcommand{\vp}{{\bm{p}}}
\newcommand{\vq}{{\bm{q}}}

\newcommand{\vu}{{\bm{u}}}
\newcommand{\vv}{{\bm{v}}}

\newcommand{\mc}{{\bm{C}}}

\newcommand{\mf}{{\bm{F}}}

\newcommand{\mk}{{\bm{K}}}

\newcommand{\matp}{{\bm{P}}}
\newcommand{\mq}{{\bm{Q}}}

\newcommand{\ms}{{\bm{S}}}

\newcommand{\matu}{{\bm{U}}}
\newcommand{\mv}{{\bm{V}}}

\newcommand{\He}{\mathrm{H}}
\newcommand{\T}{\mathrm{T}}

\tikzstyle{block} = [draw, rectangle, 
minimum height=4em, minimum width=4em]
\tikzstyle{input} = [coordinate]
\tikzstyle{output} = [coordinate]
\tikzstyle{pinstyle} = [pin edge={to-,thin,black}]

\usetikzlibrary{positioning}

\tikzset{radiation/.style={{decorate,decoration={expanding waves,angle=90,segment length=5pt}}}}

\usetikzlibrary{spy}
\usepackage{pgfplots}
\usepackage{wrapfig}
\usetikzlibrary{arrows,shapes}
\usetikzlibrary{positioning,shapes.callouts}
\usepgfplotslibrary{groupplots,dateplot}
\usetikzlibrary{patterns,shapes.arrows}
\pgfplotsset{compat=newest}

\def\BibTeX{{\rm B\kern-.05em{\sc i\kern-.025em b}\kern-.08em
		T\kern-.1667em\lower.7ex\hbox{E}\kern-.125emX}}

\usepackage[acronym,shortcuts]{glossaries}
\newacronym{DL}{DL}{downlink}
\newacronym{AP}{AP}{access point}
\newacronym{cVAE}{cVAE}{conditional variational autoencoder}
\newacronym{MSE}{MSE}{mean square error}
\newacronym{MMSE}{MMSE}{minimum mean square error}
\newacronym{WMMSE}{WMMSE}{weighted minimum mean square error}
\newacronym{SWMMSE}{SWMMSE}{stochastic WMMSE}
\newacronym{CSI}{CSI}{channel state information}
\newacronym{GNN}{GNN}{graph neural network}
\newacronym{ULA}{ULA}{uniform linear array}
\newacronym{FC}{FC}{fully connected}
\newacronym{MD-GNN}{MD-GNN}{multi-dimensional graph neural network}
\newacronym{MU-MISO}{MU-MISO}{multi-user multiple-input-single-output}
\newacronym{MISO}{MISO}{multiple-input-single-output}
\newacronym{GAT}{GAT}{graph attention network}
\newacronym{DNN}{DNN}{deep neural network}
\newacronym{MLP}{MLP}{multi-layer perceptron}
\newacronym{CNN}{CNN}{convolutional neural network}
\newacronym{SNR}{SNR}{signal-to-noise ratio}
\newacronym{OOD}{OOD}{out-of-distribution}

\tikzset{
    PlotJointSWMMSE/.style={mark=o, mark size=2pt, line width=1pt, color=blue!70!black, mark options=solid},
    PlotJointWMMSE/.style={mark=pentagon, mark size=2pt, line width=1pt, color=black, mark options=solid, dashed},
    PlotGreedySWMMSE/.style={mark=triangle, mark size=2pt, line width=1pt, color=cyan, mark options=solid},
    PlotGreedyEigenbeam/.style={mark=diamond, mark size=2pt, line width=1pt, color=olive, mark options=solid},
    PlotGNN/.style={mark=square, mark size=2pt, line width=1pt, color=orange, mark options=solid},
    PlotRandomSWMMSE/.style={mark=asterisk, mark size=2pt, line width=1pt, color=violet, mark options=solid},
    PlotNoMarkerJointSWMMSE/.style={mark=none, mark size=1.5pt, line width=1pt, color=blue!70!black, mark options=solid},
    PlotNoMarkerJointWMMSE/.style={mark=none, line width=1pt, color=black, mark options=solid, dashed},
    PlotNoMarkerGreedySWMMSE/.style={mark=none, mark size=1.5pt, line width=1pt, color=cyan, mark options=solid},
    PlotNoMarkerGreedyEigenbeam/.style={mark=none, mark size=1.5pt, line width=1pt, color=olive, mark options=solid},
    PlotNoMarkerGNN/.style={mark=none, mark size=1.5pt, line width=1pt, color=orange, mark options=solid},
    PlotNoMarkerRandomSWMMSE/.style={mark=none, mark size=1.5pt, line width=1pt, color=violet, mark options=solid}
}

%% file: intro.tex
\section{Introduction}
Indoor environments increasingly rely on dense deployments of multiple \acp{AP} to provide reliable coverage and high data rates in multi-user scenarios.
Users are commonly assigned to the \ac{AP} with the strongest received signal.
This greedy assignment can be suboptimal, particularly when users are spatially clustered.
In this case, a single \ac{AP} may end up serving all clustered users, while the remaining \acp{AP} stay idle.
To overcome this limitation, the authors in~\cite{sanjabi} formulate a sum-rate maximization problem over the assignment variables and precoding vectors. 
They solve it via alternating optimization, updating the precoding vectors via the \ac{WMMSE} algorithm and the assignment variables via a projected gradient descent step.
However, the approach in~\cite{sanjabi} assumes perfect \ac{CSI} at the \acp{AP}.
In contrast, the \ac{SWMMSE} algorithm was developed to solve the sum-rate maximization problem under statistical \ac{CSI}~\cite{razaviyayn}. 
However, this algorithm assumes a fixed pairwise transmitter-receiver association and does not optimize the assignment of users to \acp{AP}.

Because \ac{WMMSE}-based algorithms require multiple iterations to converge, \ac{DNN}-based alternatives have been proposed to reduce inference latency in precoder design. In particular, \acp{GNN} have been widely adopted for precoding in \ac{MU-MISO} systems~\cite{mu_miso_precoding_v1,mu_miso_precoding_v2}.
\acp{GNN} offer an advantage over other \ac{DNN} architectures, such as \acp{MLP} or \acp{CNN}, since they can leverage the inherent permutation equivariance property of many wireless policies.
Exploiting this property allows \acp{GNN} to avoid learning a separate mapping for every possible permutation within each index set.
A systematic approach to designing such \acp{GNN} for multiple different wireless policies is presented in~\cite{liu}.
Another advantage of \acp{GNN} is their ability to generalize to varying numbers of users, despite being trained on a single configuration~\cite{turan}.
Furthermore, an Edge-\ac{GNN} computes the precoder as an ``edge-level'' task on a bipartite antenna-user graph~\cite{rizzello}. This architecture has been shown to require fewer parameters and achieve lower inference time than vertex-based \acp{GNN}~\cite{vertex_vs_edge_gnn,rizzello}.
Yet, the standard Edge-\ac{GNN} aggregates all interfering edges with the same weight, thereby limiting its expressiveness. This is addressed in~\cite{liu} by incorporating an attention mechanism into the framework, which is shown to improve generalization across varying numbers of users. 
However, the aforementioned works still assume knowledge of perfect \ac{CSI} or a noisy estimate thereof. 
In~\cite{turan}, an attention-based Edge-\ac{GNN} is designed for statistical \ac{CSI} instead.

All \ac{GNN}-based approaches discussed above either assume a single \ac{AP} or a predetermined association between users and \acp{AP}, and thus do not address \ac{AP} assignment.
For multiple \acp{AP}, the authors in~\cite{lyu} propose a \ac{GNN} that jointly learns assignment variables and beamforming vectors. 
They argue that relaxing the integer assignment constraint introduces bias, resulting in suboptimal \ac{AP} selections.
To avoid this, they propose a Gumbel-Softmax reparameterization with a straight-through estimator to obtain a hard, integer-valued assignment.
Their \ac{GNN} employs a message-passing architecture without an attention mechanism and assumes perfect \ac{CSI}.

\emph{Contributions:}
We consider the joint sum-rate maximization problem over the assignment variables and precoding vectors, as in~\cite{sanjabi}.
We extend this setting to statistical \ac{CSI}.
In our formulation, the objective is linear in the assignment variables, which allows us to relax them to continuous values without loss of optimality.
This is in contrast to~\cite{lyu}, where the objective is non-linear in the assignment variables.
\begin{itemize}
\item
First, we present an iterative algorithm that extends the alternating optimization framework of~\cite{sanjabi} to statistical \ac{CSI}, incorporating the \ac{SWMMSE} algorithm~\cite{razaviyayn} for precoder updates.
\item
Second, we introduce a \ac{GNN}-based framework that solves the same problem in a single forward pass, thereby drastically reducing latency.
In particular, we extend the attention-based Edge-\ac{GNN} architecture from~\cite{rizzello,turan} to a multiple 
\ac{AP} scenario by adapting the underlying graph structure and update equation.
To jointly learn the assignment variables, we introduce an \ac{AP} selection head based on the standard softmax function.
\item
We empirically observe that the \ac{GNN} attains a higher sum-rate than the iterative algorithm across the tested \ac{SNR} range and generalizes to varying numbers of users with similar performance.
Both approaches outperform baselines with a fixed greedy or random \ac{AP} selection.
\end{itemize}

%% file: system.tex
\section{Preliminaries}
\subsection{System Model and Optimization Problem}
\label{sec:system_model}
We consider an indoor \ac{DL} \ac{MISO} system with $K$ \acp{AP} and $N$ single antenna users. 
Each \ac{AP} $k \in \{1,\ldots,K\}$ is equipped with $M$ antennas and each user $n \in \{1,\ldots,N\}$ is served by exactly one \ac{AP}. 
The \ac{AP} selection is denoted via the binary assignment variable $a_{k,n} \in \{0,1\}$, with $a_{k,n} = 1$ if user $n$ is served by \ac{AP} $k$, and $a_{k,n} = 0$ otherwise. 
The channel between \ac{AP} $k$ and user $n$ is denoted by $\vh_{k,n} \in \mathbb{C}^M$. 
Only statistical \ac{CSI} is assumed to be available, given by the covariance matrix $\mc_{k,n} \in \mathbb{C}^{M \times M}$, such that $\vh_{k,n} \sim \mathcal{N}_{\mathbb{C}}(\mathbf{0},\mc_{k,n})$.
The intended signal $x_n \in \mathbb{C}$ for user $n$ is precoded via its corresponding precoding vector $\vv_{k,n} \in \mathbb{C}^{M}$. 
We assume zero-mean, statistically independent signals with unit power $\mathbb{E}[|x_n|^2] = 1$.
The received signal at user $n$ for the signal transmitted at \ac{AP} $k$ is given by 
\begin{equation}
    \label{eq:signal}
    y_{k,n} = \vh_{k,n}^\He \vv_{k,n} x_{n} + \sum_{n' \neq n} \vh_{k,n}^\He \vv_{k,n'} x_{n'} + z_n,
\end{equation}
where $z_n \sim \mathcal{N}_{\mathbb{C}}(0,\sigma^2)$ denotes the complex additive white Gaussian noise.
The expected rate for user $n$ served by \ac{AP} $k$ is then given by
\begin{equation}
    \label{eq:rate}
    r_{k,n} = \mathbb{E}_{\vh_{k,n}}\left[\log_2\left(1+\frac{|\vh_{k,n}^\He \vv_{k,n}|^2}{\sigma^2 + \sum_{n' \neq n} |\vh_{k,n}^\He \vv_{k,n'}|^2}\right)\right].
\end{equation}
As in~\cite{weisser}, we assume orthogonal resource allocation across \acp{AP} and thus no inter-\ac{AP} interference.
For systems where this is not the case, the interference terms in \eqref{eq:signal} and \eqref{eq:rate} include an additional sum over $K$.

Building on the joint assignment and beamforming formulation in~\cite{sanjabi}, the sum-rate maximization problem under statistical \ac{CSI} reads as
\begin{equation}
\label{eq:OP}
\begin{aligned}
    \max_{\{a_{k,n}, \vv_{k,n}\}} \quad & \sum_{k=1}^K \sum_{n=1}^N a_{k,n} r_{k,n} \\
    \text{s.t.} \quad & a_{k,n} \in \{0,1\}, \quad \forall k,n, \\
    & \sum_{k=1}^K a_{k,n} = 1, \quad \forall n, \\
    & \sum_{n=1}^N \|\vv_{k,n}\|_2^2 \le P_\mathrm{t}, \quad \forall k.
\end{aligned}
\end{equation}
As in~\cite{sanjabi}, we do not explicitly enforce $\vv_{k,n} = \mathbf{0}$ for $a_{k,n}=0$.
It is suboptimal to allocate power to $\vv_{k,n}$ if $a_{k,n}=0$, since it only contributes to the interference.
Consequently, $\vv_{k,n} = \mathbf{0}$ naturally holds at the optimum.
For this reason, we neither have to weight the interference terms in~\eqref{eq:signal} and~\eqref{eq:rate} or the power constraint in~\eqref{eq:OP} by $a_{k,n}$ nor add a specific constraint enforcing $\vv_{k,n} = \mathbf{0}$.
As a result, our objective remains linear in the assignment variables. 

\subsection{Simulation Environment}
\label{sec:simulation}
Building upon previous work~\cite{weisser}, we evaluate our approach using the same indoor scenario.
We simulate the propagation environment of a 10\,m by 10\,m room with a height of 2.5\,m, including fixed structural elements such as walls and other objects.
Compared to~\cite{weisser}, we do not account for a variable blocking object.
The $K=2$ \acp{AP} are placed in the middle of opposite walls at a mounting height of 2.4\,m.
Each \ac{AP} is modeled by a \ac{ULA} facing towards the room center.
For later evaluations, we consider different scenarios with $M \in \{4, 8, 16, 32\}$ antennas per \ac{AP} with half-wavelength spacing.
Channel realizations are generated at a carrier frequency of 28\,GHz using the ray-tracing tool Sionna~\cite{sionna}.
Assuming that each \ac{AP} transmits with the same full power $P_\mathrm{t}$, the average \ac{SNR} of the \ac{AP}-user pair $(k,n)$ is given by
\begin{equation}
    \label{eq:SNR}
    \mathrm{SNR}_{k,n} = \frac{P_\mathrm{t}\, \mathbb{E}[\|\vh_{k,n}\|_2^2]}{M\sigma^2}.
\end{equation}
For the remainder of this work, we assume to be given statistical \ac{CSI} in the form of covariance matrices $\{\{\mc_{k,n}\}_{k=1}^K\}_{n=1}^N$.
One option for obtaining the covariance matrices is to learn the channel distribution conditioned on the user's position, as done by the \ac{cVAE} in~\cite{weisser}.
Given the position of the $n$-th user $\vp_n$ as context, the \ac{cVAE} provides a Gaussian parametrization, such that $\vh_{k,n} \mid \vp_n \sim \mathcal{N}_{\mathbb{C}}(\mathbf{0}, \mc_{k,n})$.

%% file: method.tex
\section{Method}
The optimization problem in~\eqref{eq:OP} is a mixed-integer program with a non-convex objective.
Solving it directly would require an exhaustive search over all possible assignments, which becomes computationally infeasible as $K$ and $N$ grow.
Let $\va_n = [a_{1,n}, \ldots, a_{K,n}]^\T$ denote the assignment vector of user $n$ across all \acp{AP}, and define $\mathcal{A} = \{ \va \in \{0,1\}^K : \sum_{k=1}^K a_{k} = 1\}$, such that $\va_n \in \mathcal{A}$ for all $n$.
To relax $\mathcal{A}$, we replace the binary constraint, resulting in
\begin{equation}
    \tilde{\mathcal{A}} = \left\{ \va \in \mathbb{R}_{\geq 0}^K : \sum_{k=1}^K a_{k} = 1 \right\},
\end{equation}
such that $\va_n \in \tilde{\mathcal{A}}$ for all $n$.
For fixed precoders, the objective in~\eqref{eq:OP} is linear in $\va_n$.
The optimal value of a linear objective over a polytope is attained at a vertex of that set.
Since $\tilde{\mathcal{A}}$ is the standard simplex in $\mathbb{R}^K$, its vertices are the standard basis vectors, which coincide with the elements of $\mathcal{A}$.
For fixed precoders, an optimal assignment over $\tilde{\mathcal{A}}$ is therefore binary and attains the same optimal objective value as over $\mathcal{A}$.
We thus consider~\eqref{eq:OP} over the relaxed set $\tilde{\mathcal{A}}$ in the remainder.
This relaxation and the underlying vertex argument follow from~\cite{sanjabi}.

\subsection{Stochastic WMMSE for Joint AP Selection and Precoding}
\label{sec:joint_SWMMSE}
For the joint optimization of the assignment variables and precoding vectors, an algorithm under perfect \ac{CSI} was proposed in~\cite{sanjabi}.
Under a fixed pairwise transmitter-receiver association, the \ac{SWMMSE} algorithm~\cite{razaviyayn} solves the sum-rate maximization problem under statistical \ac{CSI}.
We combine both approaches to solve the joint \ac{AP} selection and precoding sum-rate 
maximization problem under statistical \ac{CSI}.
To this end, we tackle~\eqref{eq:OP} by alternately optimizing over the precoding vectors and the relaxed assignment variables. The pseudocode is summarized in Algorithm~\ref{alg:joint_swmmse}.

As input, the algorithm requires the covariance matrices $\mc_{k,n}$ for each \ac{AP}-user pair to sample channel realizations.
Initially, the assignment of each user towards any \ac{AP} is equally weighted.
The algorithm consists of an outer loop with $L_{\mathrm{o}}$ iterations, each containing an inner loop of $L_{\mathrm{i}}$ iterations.
The inner loop is based on the \ac{SWMMSE} algorithm from~\cite{razaviyayn}, operating on a fixed \ac{AP} selection.
Note that we adapted the formulas to fit our \ac{MISO} case and optimization problem. 
Here, $u_{k,n} \in \mathbb{C}$ denotes the \ac{MMSE} receiver for user $n$ served by AP $k$.
The corresponding weight $w_{k,n} \in \mathbb{R}_{\geq 0}$ is the inverse \ac{MSE}.
Given the current precoders, both quantities are updated in closed form.
The channel statistics accumulate in $\ms_k$ and $\vb_{k,n}$.
Both are multiplied by a forgetting factor $\rho$ to suppress the influence of precoding vectors computed under assignments that were still suboptimal in earlier iterations.
The precoding vectors $\mv_k = [\vv_{k,1},\ldots,\vv_{k,N}]$ at each AP are then computed.
Here, the optimal Lagrangian multiplier $\mu_k^\star$ is obtained via a bisection search enforcing $\|\mv_k\|_\mathrm{F}^2 = P_{\mathrm{t}}$, where $\|\cdot\|_\mathrm{F}$ denotes the Frobenius norm.
After the first $L_{\mathrm{i}}$ iterations yield a well-informed precoding vector, we use an additional $L_{\mathrm{i}}$ realizations to compute a rate estimate $\hat{r}_{k,n}$ based on these precoding vectors.
Averaging over multiple channel realizations is necessary since only statistical \ac{CSI} is available.
Consequently, the averaged rate estimate allows for a more stable update of the assignment variables via projected gradient descent.
The projection onto the simplex $\tilde{\mathcal{A}}$ is done according to~\cite{duchi}. 
Note that we do not solve immediately for the globally optimal assignment variables, as this would prematurely commit to an assignment based on precoding vectors that are not yet fully optimized~\cite{sanjabi}. 
Finally, after $L_{\mathrm{o}}$ iterations, the algorithm returns the assignment variables and precoding vectors for each \ac{AP}-user combination.
\input{alg_joint_swmmse}

\subsection{GNN for Joint AP Selection and Precoding}
\label{sec:joint_GNN}
Unlike the iterative approach, a \ac{GNN} requires only a single forward pass at inference time, thereby significantly reducing latency.
Additionally, an advantage of the \ac{GNN} over other machine learning approaches stems from the structure of the sum-rate maximization problem itself.
Relabeling the user index $n$, the \ac{AP} index $k$, or the antenna index $m$ within an \ac{AP} should reorder the solution accordingly.
This property is known as permutation equivariance.
In contrast to standard \acp{DNN}, such as \acp{MLP} or \acp{CNN}, we can design \acp{GNN} to satisfy this permutation equivariance by construction.
Without it, the network would have to learn the same mapping separately for every possible labeling of users, \acp{AP}, and antennas, which inflates the sample complexity and hinders generalization to unseen numbers of users~\cite{liu}.

To this end, we extend the Edge-\ac{GAT} architecture from \cite{turan,rizzello}, first proposed as A2D-GNN in \cite{liu}.
In particular, we modify the update equation to account for multiple \acp{AP} and introduce an \ac{AP}-specific selection head.
To prepare the input data for our \ac{GNN}, namely to construct a permutation-equivariant feature set, we adapt the approaches from~\cite{rizzello,turan}.
Due to modeling the \acp{AP} as \acp{ULA}, we can leverage the Toeplitz structure of the covariance matrix by only having to consider its first row $\vc_{k,n} = \mc_{k,n}[1,:]$.
Precisely, the real and imaginary parts of $\vc_{k,n}$ are stacked and pushed through the same feature extractor $\mf_{\theta}: \mathbb{R}^{2M} \rightarrow \mathbb{R}^{2M}$ for all $k,n$.
$\mf_{\theta}$ is designed as a single \ac{FC} layer, followed by a PReLU activation function.

\begin{figure*}
    \centering
    \resizebox{0.7\textwidth}{!}{\includegraphics[]{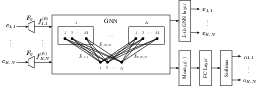}}
    \caption{Proposed GNN-based framework. A shared GNN backbone of $L-1$ layers feeds a precoder head and an AP selection head.}
    \label{fig:framework}
\end{figure*}
We utilize the \ac{MD-GNN} structure with independent and nested sets~\cite{liu}. The users constitute a flat, freely permutable set. The antennas, in contrast, are partitioned into AP-specific subsets. Permutations are unrestricted within each subset but do not extend across subsets.
Accordingly, we model the underlying network topology as a composite bipartite graph, where each \ac{AP} groups a set of antennas connected to a set of user nodes shared across all \acp{AP}.
As in \cite{turan}, features are only defined over the edges, not the nodes. 
These feature vectors are processed by a \ac{GNN} backbone of $L-1$ layers, producing shared representations for the subsequent \ac{AP} selection and precoding heads.

The input to the \ac{GNN} consists of the stacked initial feature vectors $\vf_{k,n}^{(0)} = [\vf_{k,n,1}^{(0),\T},\ldots,\vf_{k,n,M}^{(0),\T}]^{\T}$ for all $k,n$, with $\vf_{k,n,m}^{(0)} \in \mathbb{R}^2$. 
In general, we define $\vf_{k,n,m}^{(\ell)} \in \mathbb{R}^{d_{\ell}}$ as the feature vector of the $\ell$-th \ac{GNN} layer.
It corresponds to the edge between user $n$ and antenna $m$ of \ac{AP} $k$.
Each feature vector is updated as
\begin{multline}
    \label{eq:update}
    \vf_{k,n,m}^{(\ell)} = f_{\mathrm{act}} \Bigg( \matp_1^{(\ell)} \vf_{k,n,m}^{(\ell-1)} + \beta \matp_2^{(\ell)} \sum_{m' \neq m} \vf_{k,n,m'}^{(\ell-1)} \\
    + \gamma \sum_{n' \neq n} \bm{\alpha}_{k,n,n'}^{(\ell)} \odot \vu_{k,n',m}^{(\ell)} + \delta \matp_3^{(\ell)} \sum_{k' \neq k} 
    \sum_{m'=1}^{M} \vf_{k',n,m'}^{(\ell-1)} /M \Bigg).
\end{multline}
The update rule includes a linear attention mechanism, according to~\cite{liu}
\begin{align}
    &\vq_{k,n,m}^{(\ell)} = \mq^{(\ell)} \vf_{k,n,m}^{(\ell-1)} \quad
    \vk_{k,n,m}^{(\ell)} = \mk^{(\ell)} \vf_{k,n,m}^{(\ell-1)} \notag \\
    &\vu_{k,n,m}^{(\ell)} = \matu^{(\ell)} \vf_{k,n,m}^{(\ell-1)} \quad
    \bm{\alpha}_{k,n,n'}^{(\ell)} = \sum_{m=1}^M \vq_{k,n,m}^{(\ell)} \odot \vk_{k,n',m}^{(\ell)} /M.
\end{align}
The trainable weight matrices of the $\ell$-th layer are $\matp_1^{(\ell)},\, \matp_2^{(\ell)},\, \matp_3^{(\ell)}, \, \mq^{(\ell)}, \, \mk^{(\ell)}, \, \matu^{(\ell)} \in \mathbb{R}^{d_\ell \times d_{\ell-1}}$.
The scalars $\beta,\,\gamma$, and $\delta$ are hyperparameters and $\odot$ denotes the Hadamard product. 
The aggregated representation is passed through an activation function $f_\mathrm{act}(\cdot)$.
The first three terms of~\eqref{eq:update} are a direct extension of the update equation in \cite{turan}.

We introduced the fourth term to aggregate information across all other \acp{AP}, allowing the network to assess whether a user might be better served by a different \ac{AP}. 
No correspondence exists between antenna indices at different \acp{AP}, so any cross-\ac{AP} operation must remain invariant to per-\ac{AP} antenna permutations. 
The inner sum over antennas in the fourth term of~\eqref{eq:update} ensures this invariance.
No further terms are required, as we leverage the topology prior of the graph and only have to aggregate information from edges adjacent to the edge to be updated~\cite{liu}.

To obtain the precoding vectors and the assignment variables, we construct two heads acting on the shared feature representations of the $(L-1)$-th layer of the \ac{GNN}.
For the precoding head, we add another \ac{GNN} layer with the same update equation as in \eqref{eq:update}. 
Here, we omit the activation function. 
To ensure that the feature vectors of layer $L$ represent the real and imaginary parts of the corresponding precoding vector, $d_L$ is set to $2$. Consequently, we have $\vf_{k,n,m}^{(L)} = [\Re(v'_{k,n,m}), \Im(v'_{k,n,m})]^\T$. 
The precoding vector at the $k$-th \ac{AP} for the $n$-th user is then defined as $\vv'_{k,n} = [v'_{k,n,1}, \ldots, v'_{k,n,M}]^\T \in \mathbb{C}^M$. 
To enforce the power constraint, we normalize the precoding vectors, yielding the final set of precoders $\{\{\vv_{k,n}\}_{k=1}^K\}_{n=1}^N$.

For the \ac{AP} selection head, we aggregate over the antenna dimension via $\overline{\vf}_{k,n} = \sum_{m=1}^M \vf_{k,n,m} /M$, yielding feature representations that depend only on $k$ and $n$.
To map the high-dimensional vector $\overline{\vf}_{k,n} \in \mathbb{R}^{d_{L-1}}$ to a scalar $a'_{k,n}$, we apply the same \ac{FC} layer to every pair $(k,n)$.
To ensure that the assignment variables lie in $\tilde{\mathcal{A}}$, a natural choice is the softmax function, yielding $a_{k,n} = e^{a'_{k,n}} / \sum_{k'=1}^K e^{a'_{k',n}}$ for all $k,n$.
Finally, we obtain our set of assignment variables $\{\{a_{k,n}\}_{k=1}^K\}_{n=1}^N$. 
The proposed GNN-based framework is displayed in Fig.~\ref{fig:framework}.

To train the \ac{GNN}, we construct the training dataset 
$\mathcal{D}_\mathrm{train, u} = 
\{\{\{\mc_{k,n}^{(d)}\}_{k=1}^K\}_{n=1}^N\}_{d=1}^D$ for $D$ different scenarios, with uniformly sampled user positions, as detailed in Section~\ref{sec:simulation_results}.
The \ac{GNN} is trained by minimizing the negative expected sum-rate objective in~\eqref{eq:OP}, where the expectation in~\eqref{eq:rate} is approximated by averaging over $50$ sampled channel realizations.
We consider a scenario in which the number of \acp{AP} and antennas is fixed, while the \ac{SNR} and the number of users vary.
For this setting, we aim to train a single \ac{GNN} that generalizes across different numbers of users and \ac{SNR} ranges, rather than training a separate \ac{GNN} for each configuration.
For each training batch, we sample a different \ac{SNR} value to promote robustness across the \ac{SNR} range.

%% file: alg_joint_swmmse.tex
\begin{algorithm}[t]
\caption{SWMMSE for Joint AP Selection and Precoding}\label{alg:joint_swmmse}
\begin{algorithmic}[1]
\Require \hspace{-3.6pt}$
\{\{\mc_{k,n}\}^K_{k=1}\}^N_{n=1}, P_\mathrm{t}, \sigma^2, L_\mathrm{o}, L_\mathrm{i},\rho,\lambda,\varepsilon$
\State $\vv_{k,n} \sim \mathcal{N}_{\mathbb{C}}(0,1) \,\, \forall k,n$ s.t. $\|\mv_k\|_{\mathrm{F}}^2 = P_{\mathrm{t}}, \,\, \forall k$
\State $a_{k,n}=1/K, \,\, \forall k,n$
\State $\ms_k = \bm{0}, \,\, \forall k, 
\,\, \vb_{k,n} = \bm{0}, \,\, \forall k,n$
\For{$\ell_{\mathrm{o}}=1$ to $L_{\mathrm{o}}$}
\For{$\ell_{\mathrm{i}}=1$ to $L_{\mathrm{i}}$}
\State \hspace{-0.8em} $\vh_{k,n} \sim \mathcal{N}_{\mathbb{C}}(\bm{0},\mc_{k,n}), \,\, \forall k,n$
\State \hspace{-0.8em} $u_{k,n} = \frac{\vh_{k,n}^\He \vv_{k,n}}{\sigma^2 + \sum_{n'=1}^N |\vh_{k,n}^\He \vv_{k,n'}|^2}, \,\, \forall k,n$
\State \hspace{-0.8em} $w_{k,n} = (1 - u_{k,n}^* \vh_{k,n}^\He \vv_{k,n})^{-1}, \,\, \forall k,n$
\State \hspace{-0.8em} $\ms_k \leftarrow \rho \ms_k + \varepsilon \bm{I} + \sum\limits_{n=1}^N a_{k,n} w_{k,n} |u_{k,n}|^2 \vh_{k,n} \vh_{k,n}^\He, \,\, \forall k$
\State \hspace{-0.8em} $\vb_{k,n} \leftarrow \rho \vb_{k,n} + \varepsilon \vv_{k,n} + a_{k,n} u_{k,n} w_{k,n} \vh_{k,n}, \,\, \forall k,n$
\State \hspace{-0.8em} $\mv_k = (\ms_k + \mu_k^\star \bm{I})^{-1} [\vb_{k,1},\ldots,\vb_{k,N}], \,\, \forall k$ 
\EndFor
\State $\hat{r}_{k,n} = \frac{1}{L_{\mathrm{i}}}\sum_{\ell_{\mathrm{i}}=1}^{L_{\mathrm{i}}} \log_2\!\left(1 + \frac{|\vh_{k,n}^{(\ell_{\mathrm{i}}),\He} \vv_{k,n}|^2}{\sigma^2 + \sum_{n'\neq n} |\vh_{k,n}^{(\ell_{\mathrm{i}}),\He} \vv_{k,n'}|^2}\right)$
\Statex \hspace{1em} with $\vh_{k,n}^{(\ell_{\mathrm{i}})} \sim \mathcal{N}_{\mathbb{C}}(\bm{0},\mc_{k,n}), \,\, \forall k,n$
\State $\va_n \leftarrow \mathcal{P}_{\tilde{\mathcal{A}}}(\va_n + \lambda [\hat{r}_{1,n},\ldots,\hat{r}_{K,n}]^\T), \,\, \forall n$
\EndFor
\State\Return $\{\{a_{k,n},\vv_{k,n}\}^K_{k=1}\}^N_{n=1}$
\end{algorithmic}
\end{algorithm}
\vspace{-5pt}

%% file: simulations.tex
\section{Simulations}

\begin{figure*}
	\centering
    \includegraphics[]{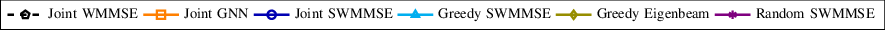}
	\subfloat[Uniformly distributed users]{
		\centering
        \includegraphics[]{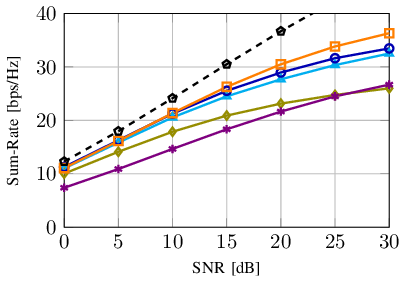}
		\label{fig:snr_rate_uniform}
	}%
	\subfloat[Spatially clustered users]{
		\centering
        \includegraphics[]{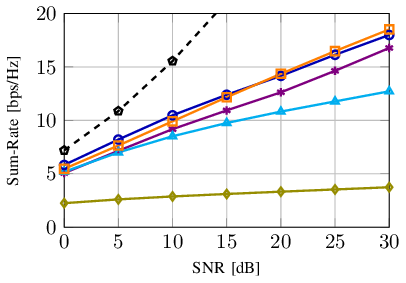}
		\label{fig:snr_rate_cluster}
	}%
	\caption{The average sum-rate over the SNR. The GNN is trained on $\mathcal{D}_{\mathrm{train,u}}$ for $K=2$ APs, $N=4$ users, and $M=16$ antennas. 
    All algorithms are tested on $\mathcal{D}_{\mathrm{test,u}}$ for (a) and $\mathcal{D}_{\mathrm{test,c}}$ for (b), with $K=2$ \acp{AP}, $N=4$ users, and $M=16$ antennas.}
	\label{fig:snr_rate}
\end{figure*}

\subsection{Baseline Methods}
In the remainder, we refer to the algorithms from Sections~\ref{sec:joint_SWMMSE} and~\ref{sec:joint_GNN} as \textit{Joint SWMMSE} and \textit{Joint GNN}, respectively.
For \textit{Joint SWMMSE}, we set $L_{\mathrm{o}}=40,L_{\mathrm{i}}=25,\rho=0.95,\lambda=0.1$, and $\varepsilon=10^{-6}$.
As an upper bound, we adapt the algorithm from~\cite{sanjabi}, which we refer to as \textit{Joint WMMSE}. 
It also jointly optimizes the precoding vectors and assignment variables, but assumes perfect \ac{CSI}, which we obtain from the ray-tracing tool described in Section~\ref{sec:simulation}.
To assess the benefit of jointly optimizing \ac{AP} assignments and precoders, we further compare against \textit{Greedy SWMMSE}, which first selects the \ac{AP}
\begin{equation}
    \label{eq:greedy}
    \hat{k}_n = \operatorname*{arg\,max}_{k \in \{1,\ldots,K\}} \ \lambda_{\max}(\mathbf{C}_{k,n})
\end{equation}
for each user $n$. Here, $\lambda_{\max}(\cdot)$ chooses the largest eigenvalue of the matrix. 
Given these fixed \ac{AP} assignments, we apply \ac{SWMMSE} to optimize the precoding vectors at each \ac{AP}.
For \textit{Random SWMMSE}, we follow the same procedure as for \textit{Greedy SWMMSE}, but replace the \ac{AP} selection with $\hat{k}_n \sim \mathcal{U}(\{1,\ldots,K\})$ for each user.

All iterative algorithms run for $1000$ iterations in total.
Since \ac{WMMSE}-based approaches require multiple iterations to converge, we additionally compare against \textit{Greedy Eigenbeam}, a closed-form baseline.
Here, the \ac{AP} selection follows equation~\eqref{eq:greedy}, and the precoding vector for each user at its assigned \ac{AP} is chosen as the scaled principal eigenvector of $\mathbf{C}_{\hat{k}_n,n}$, cf.~\cite{Ivrlac2001}.

\subsection{Simulation Results}
\label{sec:simulation_results}
For $1000$ uniformly distributed user positions, the \ac{cVAE} provides the channel covariance matrix to each \ac{AP}. 
We split these positions disjointly into $800$ training, $100$ validation, and $100$ test positions.
To train the \ac{GNN}, we generate $\mathcal{D}_{\mathrm{train,u}}$ with $D=80000$ scenarios, each containing the covariance matrices of $N=4$ user positions drawn uniformly from the training positions.
Our indoor environment consists of $K=2$ \acp{AP}, each equipped with $M=16$ antennas, unless otherwise stated, and a per-\ac{AP} maximum transmit power $P_\mathrm{t}=1$.
The \ac{GNN} backbone comprises five hidden layers, each of dimension $d_\ell = 256$ for $\ell = 1,\ldots,L-1$. 
After each layer, we employ a ReLU activation function and Group Normalization~\cite{groupnorm}.
The hyperparameters are set to $\beta = 0.1/M$, $\gamma = 0.1$, and $\delta = 0.1/K$.
The \ac{GNN} is trained for $500$ epochs with a batch size of $200$, using the Adam optimizer with an initial learning rate of $3 \times 10^{-4}$.
We use a validation dataset $\mathcal{D}_{\mathrm{val,u}}$ of $1000$ scenarios with $N=4$ users for model checkpoint selection.
For evaluation, we use separate test sets $\mathcal{D}_{\mathrm{test}}$ of $1000$ scenarios each.
Depending on the experiment, we vary the number of users and their spatial distribution across these test sets.
All algorithms are evaluated on the same test set. 

We observe that the initially relaxed assignment variables almost always converge towards binary values for \textit{Joint WMMSE}, \textit{Joint SWMMSE}, and \textit{Joint GNN}.
Only in special cases where a user is difficult to serve by either \ac{AP}, the corresponding precoding vectors approach zero and the assignment variables for that user remain ambiguous, with $\va_n \approx [0.5, 0.5]^\T$.
Moreover, we observe that when $a_{k,n}=0$, the corresponding precoding vector $\vv_{k,n}$ also approaches zero, consistent with the discussion in Section~\ref{sec:system_model}.
Regardless, for a fair comparison we obtain the hard assignment for all users via $\hat{k}_n = \arg\max_{k \in \{1,\dots,K\}} a_{k,n}$, before evaluating the rate.
The rate is averaged across the $1000$ scenarios of the corresponding test set.
\begin{figure}[t]
	\centering
    \includegraphics[]{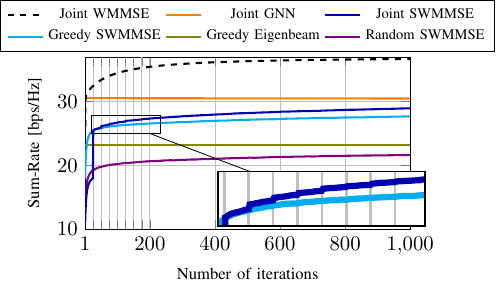}
    \vspace{-7pt}
	\captionof{figure}{The average sum-rate over the number of iterations. The GNN is trained on $\mathcal{D}_{\mathrm{train,u}}$ for $K=2$ APs, $N=4$ users, and $M=16$ antennas. All algorithms are tested on $\mathcal{D}_{\mathrm{test,u}}$ for $K=2$ APs, $N=4$ users, and $M=16$ antennas. The SNR is set to $20\,\mathrm{dB}$.}
	\label{fig:iter_rate}
	\vspace{10pt}
    
    \includegraphics[]{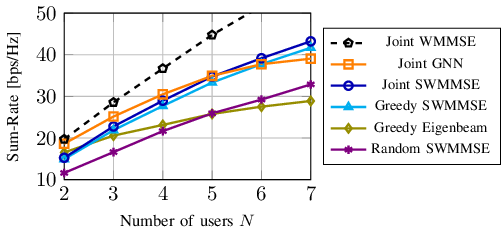}
    \vspace{-7pt}
	\captionof{figure}{The average sum-rate over the number of users $N$. The GNN is trained on $\mathcal{D}_{\mathrm{train,u}}$ for $K=2$ APs, $N=4$ users, and $M=16$ antennas. All algorithms are tested on separate datasets $\mathcal{D}_{\mathrm{test,u}}$ for $K=2$ APs, $N \in \{2,\ldots,7\}$ users, and $M=16$ antennas. The SNR is set to $20\,\mathrm{dB}$.}
	\label{fig:user_rate}
    \vspace{10pt}
    
    \includegraphics[]{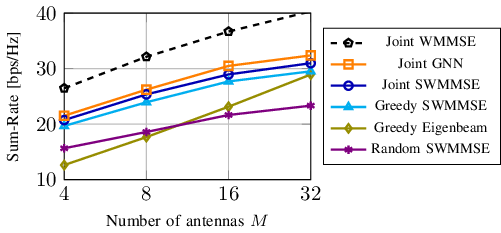}
    \vspace{-7pt}
    \captionof{figure}{The average sum-rate over the number of antennas $M$. Separate GNNs are trained on separate datasets $\mathcal{D}_{\mathrm{train,u}}$ for $K=2$ APs, $N=4$ users, and $M \in \{4,8,16,32\}$ antennas. All algorithms are tested on separate datasets $\mathcal{D}_{\mathrm{test,u}}$ for $K=2$ APs, $N=4$ users, and $M \in \{4,8,16,32\}$ antennas. The SNR is set to $20\,\mathrm{dB}$.}
    \label{fig:antenna_rate}
    \vspace{-5pt}
\end{figure}

In Fig.~\ref{fig:snr_rate_uniform}, we observe that \textit{Joint GNN} outperforms \textit{Joint SWMMSE}, especially in high-\ac{SNR} regions. 
This is likely because the \ac{GNN} learns prior information during training, which becomes particularly beneficial at high \ac{SNR}, where the interference term dominates over noise and must be resolved accurately for good performance.
Moreover, we observe that \textit{Joint SWMMSE} slightly outperforms \textit{Greedy SWMMSE}, indicating that joint optimization of the assignment variables improves performance.
Due to the uniform sampling of user positions, most assignments are trivial and thus equally well solved by a greedy assignment.
To further demonstrate this, in Fig.~\ref{fig:snr_rate_cluster} we evaluate on a test set $\mathcal{D}_{\mathrm{test,c}}$ comprising clustered users.
For each scenario, we draw one test position uniformly at random and sample the $N$ users uniformly from its $5N$ nearest test positions.
Here, a greedy \ac{AP} assignment is suboptimal, since it may result in one \ac{AP} supplying all users while the other remains idle.
Consequently, a larger gap to \textit{Joint SWMMSE} can be observed.
A random \ac{AP} selection avoids this failure mode and therefore surpasses the greedy one, but is still outperformed by \textit{Joint SWMMSE}.
The \textit{Joint GNN} still performs as well as \textit{Joint SWMMSE}, despite being evaluated on an \ac{OOD} test set.

Fig.~\ref{fig:iter_rate} shows the convergence behavior of the iterative algorithms.
For \textit{Joint SWMMSE}, the rates are evaluated at the end of each inner loop.
Dominant jumps in rate occur every $L_\mathrm{i}=25$ iterations during the initial outer iterations, due to the assignment variable update.
As the number of iterations increases, these jumps gradually diminish, leading to a smooth progression of the curve.
Since \textit{Joint GNN} and \textit{Greedy Eigenbeam} do not utilize any iterative optimization, the results are provided directly after one iteration.

As shown in Fig.~\ref{fig:user_rate}, the \textit{Joint GNN} generalizes well across different numbers of users, despite being trained only for $N=4$ users.
We see a slight performance decrease for $N=7$ users.
A similar behavior was observed for the attention-based \ac{GNN} in~\cite{liu}.
Fig.~\ref{fig:antenna_rate} shows the performance with varying numbers of antennas at the \acp{AP}.
Since $M$ is fixed in practice, we train a separate \ac{GNN} for each considered number of antennas.
The \textit{Joint GNN} outperforms \textit{Joint SWMMSE} across all configurations.
For $M=32$ antennas, \textit{Greedy Eigenbeam} becomes almost identical to \textit{Greedy SWMMSE}, as interference is inherently low given the large number of antennas relative to the $N=4$ users.

%% file: conclusion.tex
\section{Conclusion}
We proposed an iterative and a \ac{GNN}-based approach to solve the joint \ac{AP} and precoding sum-rate maximization problem under statistical \ac{CSI}.
The \ac{GNN}-based approach generalizes across users and outperforms the iterative algorithm over the \ac{SNR} range while additionally only requiring one forward pass during inference.
Both approaches outperform baselines relying on a predetermined greedy or random \ac{AP} selection.
Future work could address \acp{AP} that jointly serve a user on a shared carrier, as well as users moving along trajectories.